\documentclass{svproc}
\usepackage{hyperref}

\makeatletter
\newcommand{\printfnsymbol}[1]{%
  \textsuperscript{\@fnsymbol{#1}}%
}
\makeatother
\usepackage{cite}
\usepackage{amsmath,amssymb,amsfonts}

\usepackage{graphicx}

\usepackage{algorithm2e}
\usepackage{textcomp}
\usepackage{xcolor}
\usepackage{cite}
\usepackage[utf8]{inputenc}
\usepackage[T1]{fontenc}

\usepackage{subfig}

\usepackage{algorithmic}
\usepackage[T1]{fontenc}
\usepackage{lmodern}
\usepackage{graphicx}
\usepackage{textcomp}
\usepackage{xcolor}

\usepackage{newunicodechar}

\begin{document}
\mainmatter              
\title{Cluster Based Deep Contextual Reinforcement Learning for top-k Recommendations}
\author{{Anubha Kabra, Anu Agarwal, Anil Singh Parihar}\\
\institute{Department of Computer Science,
Delhi Technological University\\    
anubhakabradtu@gmail.com,anuag7699@gmail.com,
anil@dtu.ac.in}}

\titlerunning{Cluster Based Deep Contextual Reinforcement Learning for top-k Recommendations}  
%

\maketitle              

\begin{abstract}
Rapid advancements in the E-commerce sector over the last few decades have led to an imminent need for personalized, efficient and dynamic recommendation systems. To sufficiently cater to this need, we propose a novel method for generating top-k recommendations by creating an ensemble of clustering with reinforcement learning. We have incorporated DB Scan clustering to tackle vast item space, hence increasing the efficiency multi-fold. Moreover, by using deep contextual reinforcement learning, our proposed work leverages the user features to its full potential. With partial updates and batch updates, the model learns user patterns continuously. The Dueling Bandit based exploration provides robust exploration  as compared to the state-of-art strategies due to its adaptive nature. Detailed experiments conducted on a public dataset verify our claims about the efficiency of our technique as compared to existing techniques. \\
\end{abstract}


\section{Introduction}

\noindent With the ever-growing availability of commodities and services online, the consumer community has conveniently started shifting to online mediums for fulfilment of their needs. To capture the interest of consumers online, prompt and real time recommendation systems are needs. The current recommendation systems are non dynamic i.e: they do not update themselves with time. This discards the trend that might uplift the recommendation which is more relevant to the consumer. To keep the consumer retention intact, it is necessary that the recommendation systems update themselves from time to time to learn the changing user behaviour. These changing trends can be captured using the upcoming reinforcement learning techniques. Reinforcement Learning focuses on updating the policy in batches rather than splitting the data into training and testing. When such split is made, the policy does not generally learn anything from the testing data. Following are the challenges that the system faces with static recommendation systems : \\
\textbf{1. Ignoring Trend : }Since the system is trained on the entire training set all at once, the trends present in the consumer behaviour data get mixed up and averaged-out. This leads to non-relevant recommendation for consumers. For industry, tackling the non stationary environment is very important or it may lead to consumers loosing interest. Reinforcement learning techniques update the system frequently, hence learning the changes in the environment with time.\\
\textbf{2. Disregarding Features : }Most of the existing recommendation systems rely only on the consumer and item frequency or rating. They usually don't have a provision of utilising additional features about the consumers that may be available, that may help in recommendation. Currently, the recommendation systems rely on the narrow scope of how frequently a consumer or item is seen, and expilcit user feedback in terms of ratings \cite{harper2015movielens}. With reinforcement based learning, these problems can be curbed by using contextual RL, which  considers all features available in the data set before making valid recommendations.\\
\textbf{3. No accommodation for new consumers : }Due to the training and testing steps being discrete, if a new user joins the system, the entire system needs to be retrained. This is a very cumbersome process. Reinforcement learning based strategies can easily handle this situation by updating the system with each new trial. So, if a new user joins the system, the agent will learn about the user with time without re-training.\\
\noindent This paper focuses on exploration using dueling bandit gradient descent for reinforcement learning based recommendation strategy. The proposed work also narrows the item-set to be considered for recommendation by using DB Scan clustering. The novelty lies in combining both strategies to overcome the drawbacks of current recommendation strategies and make it fit for continuously changing online environment. The combined model meets the industry standards due to its swift working, while exhibiting recommendations highly applicable to users.

\begin{figure}[!h]
\centering
\includegraphics[width=0.2\columnwidth]{"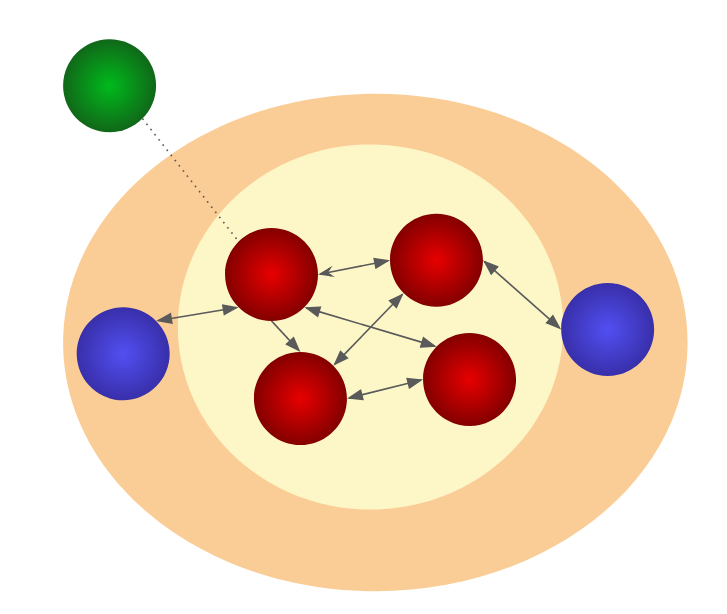"}
\caption{{DB SCAN}}
\label{fig:db}
\end{figure}

\section{Related Work}
Existing Recommendation Systems can be broadly divided into major categories.
The first category consists of the traditional recommendation algorithms which are either popularity based or collaborative. Popularity based systems always recommend the same items at given time, irrespective of the user. On the other hand, collaborative-Filtering (CF) based systems \cite{das2007google, marlin2004multiple} look for other users which are similar to the given user and recommend him/her the items already purchased by them. It is still one of  the most popular and successful techniques. Content-based filtering \cite{mooney2000content, ijntema2010ontology, ijntema2010ontology} forms another category of recommenders which generates recommendations based on items already purchased by the user. However, this technique generally doesn't work well for platforms serving items of multiple categories as this method does not generalize well across different categories. Yet another traditional algorithm is Hybrid Filtering \cite{de2012chatter, liu2010personalized} which uses a combination of above two techniques. Matrix Factorization \cite{koren2009matrix} extends CF as it attempts to identify latent features from user-ratings data by expressing the user-item interaction matrix as a product of multiple lower dimensional matrices. Further attempts \cite{zhang2019deep} have been made to incorporate the use of neural networks with the aforementioned techniques to improve the results.

The second major and relatively modern category includes recommendation systems which use reinforcement learning \cite{mahmood2009improving, taghipour2007usage, taghipour2008hybrid} and are inherently dynamic in nature.The recommendation scenario maps well to a usual Reinforcement Learning problem where an agent learns to behave in an environment via a series of sequential observations. These sequential observations can be mapped to the user-item interaction logs generated over time. Further, this task has been modelled by some as Markov Decision Problem \cite{rojanavasu2005new, lu2016partially} and others as a Multi-Armed Bandit problem \cite{li2010contextual}.
Attempts have also been made to incorporate session context into the system to provide more relevant recommendations \cite{zeng2016online, wang2016learning, cortes2018adapting}.
Our proposed work combines contextual Multi armed bandits strategy with DB-SCAN clustering Dueling Bandits exploration strategy to provide highly personalised recommendations in a more efficient manner.
\section{Methodology }

\subsection{DB Scan Clustering}
\noindent DB scan is a clustering algorithm that is density-based. The distance between points is used to find the neighbors (points withing close vicinity). This measurement is generally done on the basis of Euclidean distance. Furthermore, it specifies the out-of-bound points and low-density regions. DB scan has two parameters : 
\textbf{minPts: } For the formation of a cluster, least number of points to be considered.
\textbf{eps:} The distance between two points to be called neighbours. This distance measures the closeness.
\noindent DB scan corresponds well with the real-life data. Due to the unpredictable structure in such data with a lot of noise, DB scan algorithm can fit these irregularities well, which is generally denied in other clustering algorithms. Eg: K Means clustering generally fails when the data is not spherical in nature, whereas DB Scan can capture this.
\begin{figure*}[]
\centering
\includegraphics[width=\textwidth, height =5cm]{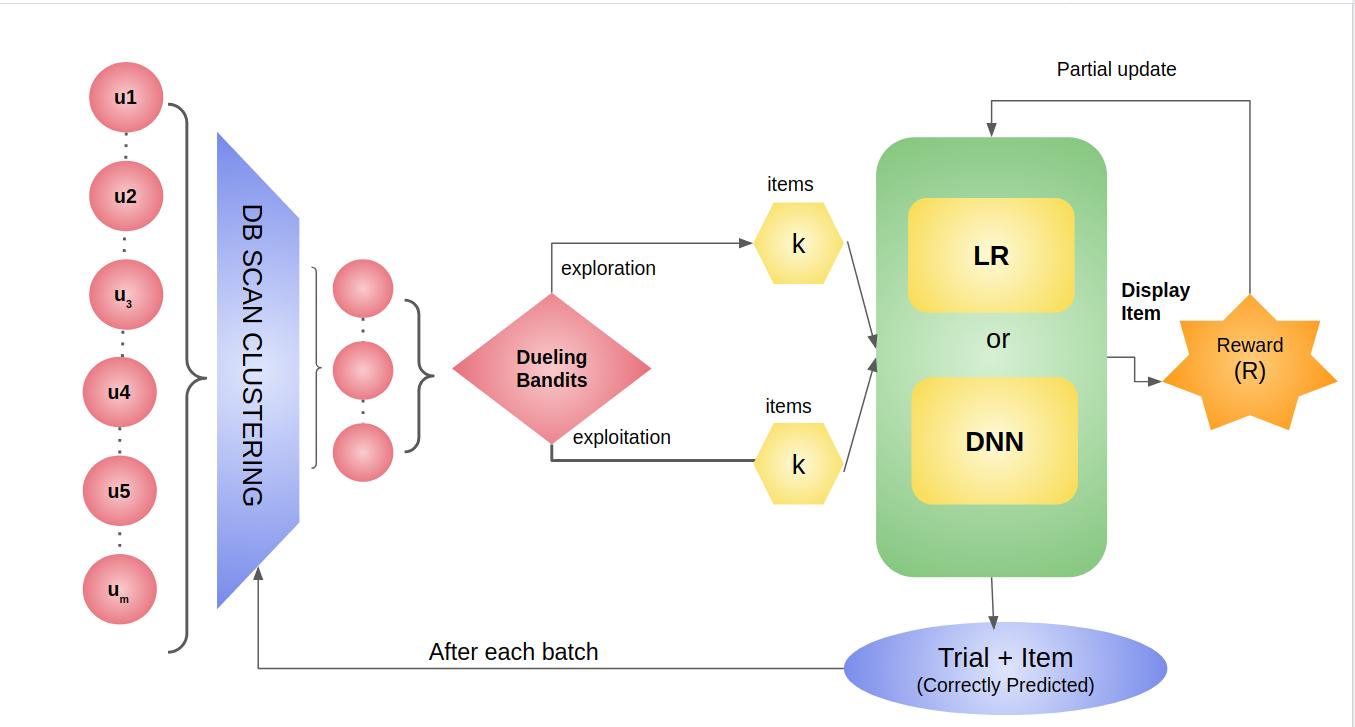}
\caption{Proposed Model Pipeline}
\label{fig:pipeline}
\end{figure*}
Figure \ref{fig:db} shows the Red points to be the core points which satisfy the minPts criteria and are in the eps vicinity. The Blue points satisfy the eps distance but exceed the minPts criteria and hence are not included in the cluster (Yellow Area). The Green Point does not satisfy any of these conditions, and hence is referred to as noise.\\

\begin{algorithm}[H]

\label{alg:algo1}
\SetAlgoLined

\textbf{Initialization}\\
D : dataset\\
$p_u$ : Un-visited Points in the dataset\\
$p_v$ : Visited points in dataset\\
C : cluster index

 \For{non-visited $p_u$ in D}{
        $p_{u_i}$ changes to $p_{v_i}$\\
        N : neighbours($p_i$,eps)\\
        \uIf{ N $>$ minPts}{
            N = N $U$ N'
        }
        \uIf{found p' does not belong to any cluster}{
            p' $\in$ C 
            }
    }    
\caption{DB Scan Algorithm (eps, minPts)}
\end{algorithm}

\subsection{Dueling Bandit Gradient Descent}
Dueling Bandit Gradient Descent is an adaptive exploration strategy unlike epsilon greedy, UCB, TS etc. It attempts to present the best recommendations to the user by combining exploration and exploitation in appropriate proportions.
As the name suggests, in this technique, two models work simultaneously. First model $M$ takes decision based on patterns learnt from the data whereas the second model $M'$ is exploratory in nature. For every row in the dataset, both models generate a list of k items each. These lists are then combined by the reinforcement learning agent using probabilistic interleaving \cite{grotov2016online} to generate a final list of k items. This combined list is presented to the user and partial feedback is obtained for each item on the list. If more items from the list generated by $M'$ get positive, then the parameters ($P$) of model $M$ are updated to incline it towards $M'$ as shown in Equation \ref{eqn:dbgd}, else the model is kept unchanged.
\begin{equation}
\label{eqn:dbgd}
    P_{new} = P_{old} + \beta P'
\end{equation}

Where other exploration strategies tend to lower the performance of the recommender in the long run because of being predominantly random in nature, this technique strikes a good balance between exploration and exploitation and adapts as and when required.

\subsection{Model Pipeline}

\noindent \textbf{STEP 1 : Clustering}
Due to the vast number of applications which can be recommended to user, DB Scan clustering can help reduce the number of items to be taken into consideration before recommendation. The initial batch size does not implement DB Scan clustering. Collection of correctly predicted user context and item is done.Clusters of items are made using context features of the first batch. For the next batch, when a trial sample is passed through this clustering mechanism, it first clusters applications relatable to the user and its features, from which best k items are selected to be shown to user with that particular features. The re-clustering is done after every batch is processes with the building collection of trail and corresponding correctly predicted arm. This technique works very well to improve efficiency since the pool of applications considered is not the entire item-set but a smaller chunk of it. Moreover it helps to understand the user behaviour with applications better.

\textbf{STEP 2: Exploration or Exploitation}
After clustering, decision is made on whether to pursue exploration(selecting an item for user with some form of randomised way) or exploitation (selecting best item for the user) however this is not as straightforward as epsilon-greedy. The strategy used for this decision is Dueling Bandits Gradient Descent which generates a list combined from both exploration and exploitation models. The exploit network adapts itself further in the process if required based on the user feedback.

\textbf{STEP 3: Reward Formulation}
The k selected items are passed through the base model to predict if user has chosen any one of them or not.There are two base models experimented with. First one is a basic Logistic Regression(LR) based model while the other is a Deep Neural Network(DNN). DNN gives higher performance than LR. If any of the k items matches with the actually selected item, the reward is 1, else the reward is 0.This reward, along with the trial features of that particular user is fed back into the model for continuous learning.

\textbf{STEP 4: Partial Update}

The model is updated in random batches on default size 1000 selected from all the trials experienced till now. The trial and reward information is stored over time in memory. Random mini batch of 1000 from this set after every 5000 trials and updated back to the base model. This helps to keep the trend of the market intact and learn a pattern similarity between context features of different users.

\section{Experimentation}

\subsection{Dataset}
The dataset used consists of user and mobile application interaction logs. The Data is collected from TalkingData SDK \footnote{https://www.kaggle.com/c/talkingdata-mobile-user-demographics/data} integrated within mobile apps TalkingData serves under the service term between TalkingData and mobile app developers.Here the apps are considered as items to be recommended and user and their information as the context used for the same. The categorical features are one hot encoded and continuous features are scaled before being given as input.

\subsection{Baselines}

\subsubsection{Static Techniques}
\hspace{\columnwidth}
\\
\textit{•LR: } Logistic Regression is widely used in the industry as a baseline method due to its easy implementation and high efficiency. It is implemented using Keras. 

\textit{•FM: } Factorization Machines is a feature handy recommendation system which takes in all contextual features and predicts clicks/non-click.

\textit{•W\&D: } Wide \& Deep combines dependent learning using\,textbf{wide} : a normal representation of the features and \textbf{deep}: embedding representation of feature
to predict the click label.
\begin{table}[ht]
\scriptsize
\centering
\caption{Abbreviation and Code Number for each policy .Proposed policies are given in \textbf{bold}. }
\label{tab:code}
\begin{tabular}{|l|l|l|}
\hline
Code & Abbreviation                      & Policy
\\ \hline
1    & LR                                & Logistic Regression                           \\
2    & W\&D                              & Wide and Deep                                 \\
3    & FM                                & Factorization Machines                        \\
4    & B-UCB                             & Bootstrapped Upper Confidence Bound           \\
5    & B-TS                              & Bootstrapped Thompson Sampling                \\
6    & E-Greedy                          & Epsilon Greedy                                \\
7    & FEE                               & First Explore then Exploit                    \\
8    & AE                                & Active Explorer                               \\
9    & \textbf{DB+LR}   & \textbf{Dueling Bandits+LR}  \\
10   & \textbf{DB +DNN} & \textbf{Dueling Bandits+DNN}\\
11  & \textbf{DB\_Scan + DB +DNN} & \textbf{DB Scan+ Dueling Bandits+DNN}\\
\hline
\end{tabular}
\end{table}


\subsubsection{Dynamic Techniques}
\hspace{\columnwidth}

\textit{•Contextual Bandits}
These are Reinforcement Learning techniques with base classifier as Logistic Regression. The following exploration strategies are explored:

\noindent\textit{•Epsilon Greedy :}The idea here is to explore with certain fixed possibility. The exploration threshold is generally kept very small. Here default is set to 0.2. This has many variations for improved exploration.

\noindent\textit{•First explore then exploit :} The exploration in this case is done initially, and later only exploitation is taken into consideration. The system assumes that initial exploration is good enough for system to learn.

\noindent\textit{•Bootstrapped UCB \cite{cortes2018adapting} :} UCB works on upper confidence bound strategy for exploration.The bootstrapping ensures small pool of trails to be taken at a single time.

\noindent\textit{•Bootstrapped TS \cite{cortes2018adapting} :} Thompson sampling creates distribution for each item, which changes according to the reward updation for the selected trial. This strategy involves forming a prior distribution for each arm. 

\noindent\textit{•Active Explorer\cite{cortes2018adapting} :} This strategy focuses on weighted label based probabilities.

\subsection{Evaluation Metrics}

\textbf{ CTR } : Click behaviour can be used to analyze how well a recommendation system is performing. CTR helps to capture the user click behaviour.

\begin{equation}
CTR_i =\frac{Total\,number\,of\,clicks}{Total\,number\,of\,occurrences}
\end{equation}

The average of CTR values of all the items is as follows:
\begin{equation}
CTR =\frac{\sum_{i=0}^{N} CTR_i}{N}
\end{equation}

\textbf{Precision@k} : Precision@k for top-k recommendations is calculated as follows:
\begin{equation}
Precision@k =\frac{number\,of\,clicks\,in\,top-k\,recommended\,items}{k}
\end{equation}

\begin{figure}[!h]
\centering
\includegraphics[width=0.40\columnwidth]{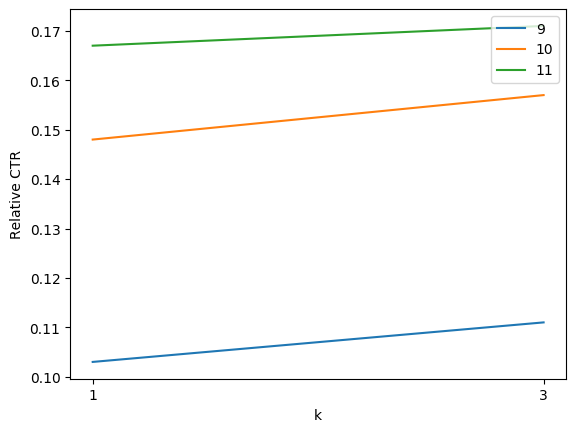}
\caption{{Comparative analysis of proposed methods for 1 and 3 item recommendations}}
\vspace{6pt}
\label{fig:comparison}
\end{figure}

\subsection{Experimentation Settings}
Three simulations were used for experimentation as shown in Table \ref{tab:code}. The techniques \textbf{9} and \textbf{10} use different base model.Technique \textbf{9} has used Logistic Regression to simulate \cite{cortes2018adapting} strategy. \textbf{10} uses Deep Neural Network(DNN) as the base model. Since DNN gives significantly better results, technique \textbf{11} uses DNN as base model with DB Scan Clustering. Experiments have been conducted for top-1 and top-3 recommendations. While static strategies have a train-test split of 70:30. The dynamic techniques rely on learning throughout the process and no split has been done. To evaluate the results however, 30\% of the data is considered, which is kept identical for all experiments.

\begin{figure}
\centering
    \subfloat[$k = 1$]{\includegraphics[width=0.45\columnwidth]{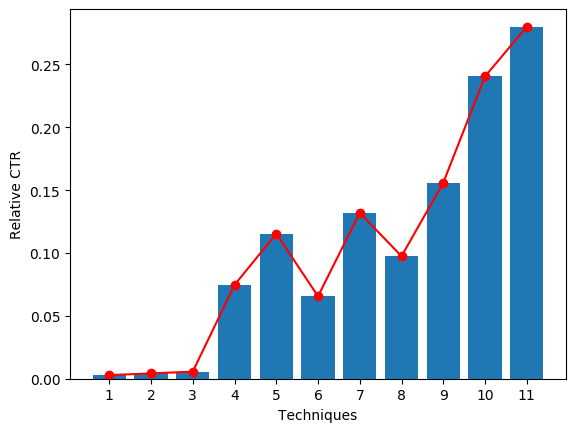}}
    \subfloat[$k = 3$]{\includegraphics[width=0.45\columnwidth]{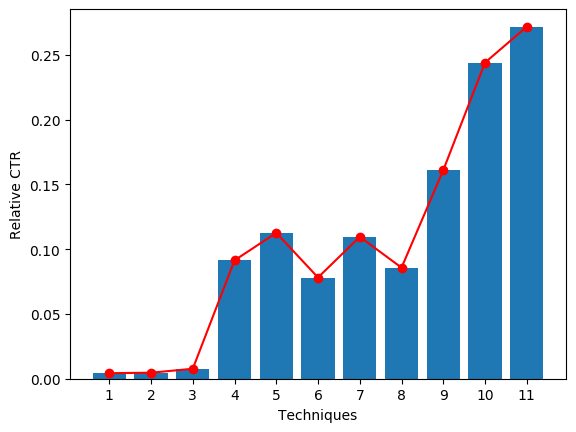}}
    
    \caption{Relative CTR trends for different values of $k$}
    \label{fig:rctr-trends}
\end{figure}

\section{Results and Analysis}


\begin{table}[ht]
\footnotesize
\centering
\caption{Result for each policy mentioned with its respective CTR and Accuracy values . }
\label{tab:results}
\begin{tabular}{|l|l|l|l|l|}
\hline
Abbreviation                        & \multicolumn{2}{l|}{CTR}                   & \multicolumn{2}{l|}{Precision@k}                                 \\ \hline
                                    & k=1\hspace{0.7cm}                              & k=3\hspace{0.7cm}    & k=1\hspace{0.5cm}                            & k=3\hspace{0.5cm}                            \\ \cline{2-5} LR                                  & 0.0022                           & 0.0042 & 0.28                           & 0.28                           \\
W\&D                                & 0.0032                           & 0.0010 & 0.33                           & 0.36                           \\
FM                                  & 0.0036                           & 0.0051 & 0.35                           & 0.32                           \\
B-UCB                               & 0.0243                           & 0.0276 & 0.54                           & 0.59                           \\
B-TS                                & 0.0243                           & 0.0276 & 0.67                           & 0.70                           \\
E-Greedy                            & 0.0288                           & 0.0323 & 0.44                           & 0.48                           \\
FEE                                 & 0.0294                           & 0.0388 & 0.69                           & 0.68                           \\
AE                                  & 0.0285                           & 0.3117 & 0.48                           & 0.49                           \\ \hline
\textbf{DB+LR}     & \textbf{0.0432} & \textbf{0.0566} & \textbf{0.75} & \textbf{0.77} \\
\textbf{DB +DNN}   & \textbf{0.0965} &\textbf{0.0972} & \textbf{0.82} & \textbf{0.85} \\
\textbf{DB\_Scan + DB + DNN} & \textbf{0.1342} & \textbf{0.0157} & \textbf{0.89} & \textbf{0.92}\\
\hline
\end{tabular}
\end{table}

It can be clearly deduced from the CTR and Precision@k values mentioned in Table \ref{tab:results} that the methods which rely on static (one-time) model training such as LR, W\&D and FM, perform poorly in comparison with the rest of the reinforcement learning based techniques. Thus, it can be concluded that dynamic model training is pivotal when it comes to applications such as recommender systems, because the user's underlying behaviour pattern keeps changing with time.Further, on comparing the CTR values for different exploration strategies, it can be seen that the adaptive techniques such as Active Explorer perform better than purely random exploration strategies like $\epsilon$-greedy. However, the best performance is given by Dueling Bandit Gradient Descent strategy as it diversifies the network and inclines it towards exploration if favourable user feedback is obtained. The relative CTR graphs in Figure \ref{fig:rctr-trends} for different values of k show that in-spite of a slight decrease in the overall values from k=1 to k=3, the general trend still remains the same. Non-RL based techniques show just about no change in results with changing k values. Since RL based techniques incorporate feedback for every recommended item, the change in values is comparatively more significant for them.Figure \ref{fig:comparison} shows a direct performance comparison between different variations of deep contextual bandits proposed in this paper. Even though DB + DNN seems to be the best strategy as it give the highest values throughout, DB\_SCAN + DB + DNN is the most efficient and practical alternative. Inclusion of clustering straightaway reduces the number of comparisons by a factor of $x/N$ where $x$ is the average size of a cluster.

\section{Conclusion}
The proposed work focuses on improving recommendations of items to users in real time scenario. Using DB Scan clustering on the data reduces the processing cost making the recommender efficient. Moreover, Reinforcement Learning techniques with Dueling Bandit Gradient Descent exploration and partial feedback tends to learn the user pattern and adapts to it quickly. Dueling Bandit exploration gives this work an edge as it performs adaptive exploration with lesser number of parameters requiring manual tuning. This leads to relevant item suggestions to users without cumbersome retraining. For future work, various other clustering methods could be experimented with for item space reduction. Ensemble of exploration strategies can also be tried.

\bibliographystyle{abbrv}
\bibliography{references}
\end{document}